\documentclass[]{aa} 

\usepackage{graphicx}
\usepackage{txfonts}
\usepackage{ulem}
\begin{document} 

   \title{Dust in the wind of outbursting young stars}

   \author{Kundan Kadam\inst{1}\thanks{email: kundan.kadam@oeaw.ac.at},
          Eduard Vorobyov\inst{2,3},
          Peter Woitke\inst{1},
          \and
          Manuel Güdel\inst{2}
          }

   \institute{Space Research Institute, Austrian Academy of Sciences, Schmiedlstrasse 6, A-8042 Graz, Austria
   \and
   Department of Astrophysics, The University of Vienna, Türkenschanzstrasse 17, 1180 Vienna, Austria
   \and
   Research Institute of Physics, Southern Federal University, Rostov-on-Don, 344090 Russia 
             }

   \date{Received February 4, 2025; accepted March 27, 2025}

 
  \abstract
   {Young stellar objects (YSOs) have been observed to undergo powerful accretion events known as FU Orionis outbursts (FUors). 
   These types of episodic accretion events are now considered to be commonplace during low-mass star formation, wherein  accretion onto the protostar occurs through a surrounding centrifugal disk. 
   Increasing evidence suggests that the magnetic disk winds are crucial for driving disk accretion, as they carry both mass and momentum away from the disk.
   }
   {
   We aim to investigate the phenomenon of the ejection of magnetic disk winds during episodic accretion, with a focus on the dust contained within these winds.
   }
   {
   We conducted magnetohydrodynamic (MHD) simulations of the formation and evolution of a protoplanetary disk (PPD) in the thin-disk limit. 
   We included the evolution of dust with two populations and a realistic prescription for viscosity during outbursts, which depends on the local thermal ionization fraction.   
   The disk evolves with the concurrent action of viscosity, self-gravity, and magnetic disk winds.
   }
   {
   The simulated disk exhibits outbursting behavior in the early stages, with 
   the duration and frequency of the bursts, their rise times, and brightness amplitudes resembling the properties observed for FUors. 
   {We find that during the outbursts, the winds are over an order of magnitude more dusty, as compared to in quiescence.
   However, despite this increased dust content, the winds are still dust-depleted as the dust-to-gas ratio is about an order of magnitude lower than the canonical interstellar value of 0.01. 
   The results of our numerical experiments are in general agreement with the available observational findings, shedding light on the mechanism behind the production of dusty winds during outbursting events in YSOs.}
   }
   {}

   \maketitle
%

\section{Introduction}

Episodic accretion is considered to be an integral part of low-mass star formation \citep{Audard+14,Fischer+23}.
As a young star accretes matter via the surrounding disk, the system suddenly gets brighter by several magnitudes and remains in this luminous state for decades or even centuries.
These eruptions are named after the prototype FU Orionis system \citep{Herbig77, Hartmann-Kenyon96}.
These are essentially enhanced accretion events, wherein the disk accretion rate rapidly increases,  typically from $\sim 10^{-7}$ to $10^{-4}$ $M_{\odot} {\rm yr^{-1}}$.
Such large and powerful outbursts have significant impact on the mass budget, 
disk chemistry, ice lines, dust evolution, mineralogy, and dynamics, especially in the innermost few astronomical units (au) of the disk.
Although a number of phenomena could feasibly explain such outbursts, the mechanism behind episodic accretion remains poorly constrained \citep{BB92,Bell-Lin94, Armitage+01, VB05,Bae+13}.

The disk that surrounds a protostar is traditionally considered to be evolving viscously \citep{SS73}, with magnetorotational instability (MRI) providing the necessary turbulent viscosity \citep{Balbus-Hawley91,Turner+14}.
However, several lines of reasoning indicate that viscosity alone is insufficient to drive the evolution of observed disks \citep{Pinte+16,Dullemond+18,Franceschi+23}.
It has been suggested that magnetic disk winds play a significant role in the angular momentum transport \citep{Bai16,Pudritz19,Lesur+22,Pascucci+23}.
The disk winds are distinct from the highly collimated jets that are often associated with star formation.
The winds are essentially magnetocentrifugal in nature and are launched from a large extent of the disk \citep{Blandford-Payne82}. 
These winds are observed as rotating, conical molecular outflows in resolved sources \citep{Gudel+18,Lee+21,Launhardt+23} or a low-velocity component (LVC) in high-spectral-resolution observations of forbidden gas lines \citep{Natta+14, McGinnis+18}; however,  it is, in fact, observationally challenging to distinguish them from photoevaporative winds \citep{Alexander+14, Rab+23}.
However, unlike photoevaporative winds, the magnetic winds carry not only mass but also angular momentum from the disk. 
The effects of magnetic winds on the disk evolution are significant, as the outflow rate is of the order of accretion rate onto the central star \citep{Watson+16, Tabone+20}.

When it comes to the nature of disk winds during episodic accretion, our understanding is incomplete due to both observational limitations and the complexity of physics involved.
Some protostellar jets are observed to show series of shock fronts or knotty structures, which are thought to be causally related to the increased accretion during episodic accretion \citep{Arce+07,Vorobyov+Knotty18}.
Although the data are sparse, the evidence suggests that the eruptions are also accompanied by enhanced extended ejections, wherein the mass-loss rate of the outflow increases up to an order of magnitude \citep{Calvet+93,Audard+14, Milliner+19}.  
Shearing box MHD simulations suggest that this may be because of the associated increase in the FUV luminosity of the central object \citep{Bai13, Bai+16}. 
The ionization layer thickness at the disk surface increases with the incident FUV radiation, resulting in an increase in the wind strength (both in terms of mass and angular momentum loss), due to better coupling with the magnetic field.
{Although it is hard to differentiate outflows from photoevaporative and magnetic winds, both observationally and theoretically, in reality, both of these may coexist as ``magneto-thermal disk wind" \citep{Clarke+05,Bai+16}.

In terms of the dust content, similar to photoevaporative winds, the magnetic disk winds preferentially remove gas and are thus significantly dust depleted \citep{Suzuki+10,Gorti+15}.
Some of the near-infrared (NIR) interferometric observations of T Tauri stars can be best explained by models of sub-micron-sized dust well above the disk midplane \citep{Labdon+23}. 
Preliminary observations of spatially resolved disk winds with HST and JWST also point towards their ability transport small dust over large distances \citep{Duchene24eas}.
Another indicator of dust content in the vicinity of T Tauri stars comes from X-ray observations, which show two component spectra, wherein the photoelectric absorption of the harder component suggests presence of dust depleted gas \citep{Gudel+07,Gudel08}.
Although it is thought that sub-micron sized dust dominates the outflows, the distribution of dust within disk winds is not well characterized. 
In the case of episodic accretion, the situation becomes more complex, for instance, because of short- and long-term time variability, projection effects, or interaction with surroundings of the infalling envelope \citep{Clarke+05,Kospal+08}.
However, recent observations show evidence of fading in young eruptive stars dominated by an increase in circumstellar extinction, which can be interpreted as presence of dusty winds \citep{Nagy+23}. In addition, deep fadings in the brightness of UX~Ori-type stars can be explained by eclipses caused by dusty winds with non-axisymmetric distribution or structural inhomogeneity} \citep{Shulman2022,Grinin+23}.

In this paper, we use a sophisticated and comprehensive global model of protoplanetary disk, which contains all the necessary ingredients to investigate the phenomenon of dust ejection during episodic accretion. 
In Sect.~\ref{sec:model}, we describe the key components of this model, which are relevant to understanding the subject at hand. 
In Sect.~\ref{sec:results}, we elaborate on our results and our findings are summarized in Sect.~\ref{sec:conclusions}.

\section{Model description}
\label{sec:model}

In this paper, we employ the modeling tool set described in \citep{Kadam+25a} and in the interest of brevity,  we summarize the most salient parts here.
The Eulerian magnetohydrodynamic (MHD) code that is used for modeling a protoplanetary disk (PPD) is called Formation and Evolution Of a Star And its circumstellar Disc (FEOSAD) and the relevant version of this code is described in detail in \cite{VMHD20}.
The following description focuses on dust evolution, outburst physics, and magnetic disk winds, as these components are most relevant for understanding the results (see \citet[][]{Kadam+25a} for more details).
The simulations are conducted in the thin disc limit, wherein the evolution of both gas and dust is considered.
The simulations start with the gravitational collapse phase of a starless cloud core, with the initial surface density and rotational profiles consistent with an axisymmetric core collapse \citep{Basu97}. 
The equations of continuity, momentum conservation, and energy transport are solved for the gas component, while the evolution of the magnetic field is considered in the ideal MHD approximation.  

The dust in our model consists of two components; submicron-sized small dust, which is fully coupled with the gas, while the grown dust component drifts with respect to the gas.
The grown dust exerts a back reaction onto the gas and evolves within the Epstein regime.
The small dust can turn into grown dust and vice versa, depending on the dust growth rate and the dynamically varying maximum size of dust grains \citep{Vorobyov+22}. 
The conversion rate between the dust components is derived based on the assumption that each of the two dust populations have a power-law function that is continuous with a power law index of $p=-3.5$, and both the minimum size of dust grains and dust mass stay constant  during dust growth. 
If the maximum size of dust grains is reduced, for example, during an outburst due to increased turbulence and temperature, the same governing equations convert grown dust into small dust. Dust growth is limited by dust radial drift, considered self consistently via the solution of dust dynamics equations for grown dust, and by the fragmentation barrier, with a fixed fragmentation velocity of 3.0~m~s$^{-1}$.

The magnetorotational instability (MRI) is considered to be the primary driver of disk viscosity. 
{The numerical viscosity is negligible in comparison to the physical one owing to the use of the third-order accurate piecewise-parabolic interpolation scheme \citep{Colella1984}.}
The disk is allowed to exhibit layered accretion \citep{Gammie96}, as the viscous evolution is considered with an adaptive \cite{SS73} $\alpha$ prescription. 
The thickness of the MRI-active layer is calculated from the ionization fraction, with the assumption that Ohmic diffusivity dominates the disk midplane. 
The ionization fraction ($x_{\rm ion}$) is obtained by solving the detailed ionization balance equation, which considers collisional ionization and recombination on dust grains as well as due to radiative processes \citep{DK14,VMHD20}. 
The disk self-gravity takes into account contribution from both the gas and dust components and the gravitational torque forms a dominant source of mass and angular momentum transport in the early stages \citep{VB09}.

The outbursts occurring in our simulations resemble the FU Orionis eruptions and are investigated in detail in a series of studies \citep{Kadam+20,Vorobyov+20,Kadam+22}.
These outbursts are termed MRI-bursts, since the underlying mechanism relates to sudden ionization of alkali metals in the inner disk due to viscous heating and successive activation of the MRI in a magnetically dead zone.
{The thermal ionization considers the contribution from Potassium only, as it dominates other alkali metals and is sufficient for assuming MRI activity \citep{DK14,VMHD20}.} 
The associated value of the $\alpha$-parameter during the bursts is assumed to be 0.1, which is motivated by 3D radiation magnetohydrodynamic simulations for FUor disk \citep{Zhu+20}.
The disk is considered outbursting above a characteristic threshold value of thermal ionization fraction, $x_{\rm th} > 10^{-10}$ \citep[][]{Kadam+25a}. 
In the non-outbursting state, the disk can have a value of $\alpha=10^{-3}$, which is consistent with canonical values of protoplanetary disks.
In addition, a magnetically dead zone is modeled as weighted average of surface densities of MRI active and dead layers, so that the dead zone corresponds to a lower effective $\alpha$ \citep{Bae+14, VMHD20}. 

The disk winds are magnetocentrifugal in nature and our model \citep[][]{Kadam+25a} is based on insights gained from shearing box simulations \citep{Bai13}.
The modeled winds exhibit the extended, conical (non-collimated), slow (low velocity component, LVC), and rotating molecular outflows observed in YSOs.
The winds depend on the local properties of the disk (magnetic $\beta$ and gas surface density), as well as its global properties (distance from the central star, stellar mass and luminosity).
The winds remove the gas mass and angular momentum from the entire extent of the centrifugal disk. 
The grown dust typically settles to the disk midplane and the winds originate from the disk surface several scale heights above the midplane. 
Hence, the winds are assumed to entrain and remove only the small dust, in proportion to the low local dust-to-gas ratio.
{Such an entrainment of sub-micron-sized dust particles by disk winds is supported by MHD simulations \citep{Tsukamoto+21,Bino+22,Rodenkirch-Dullemond22}, as well as observations \citep{Labdon+23,Duchene24eas}.} 
During an outburst, the Far-UV (FUV) luminosity of the system increases and the increase in wind strength due to the associated increase in the FUV penetration depth is taken into account.

\section{Results}
\label{sec:results}

\subsection{Long-term episodic accretion}

\begin{figure}
\centering
  \includegraphics[width=0.45\textwidth]{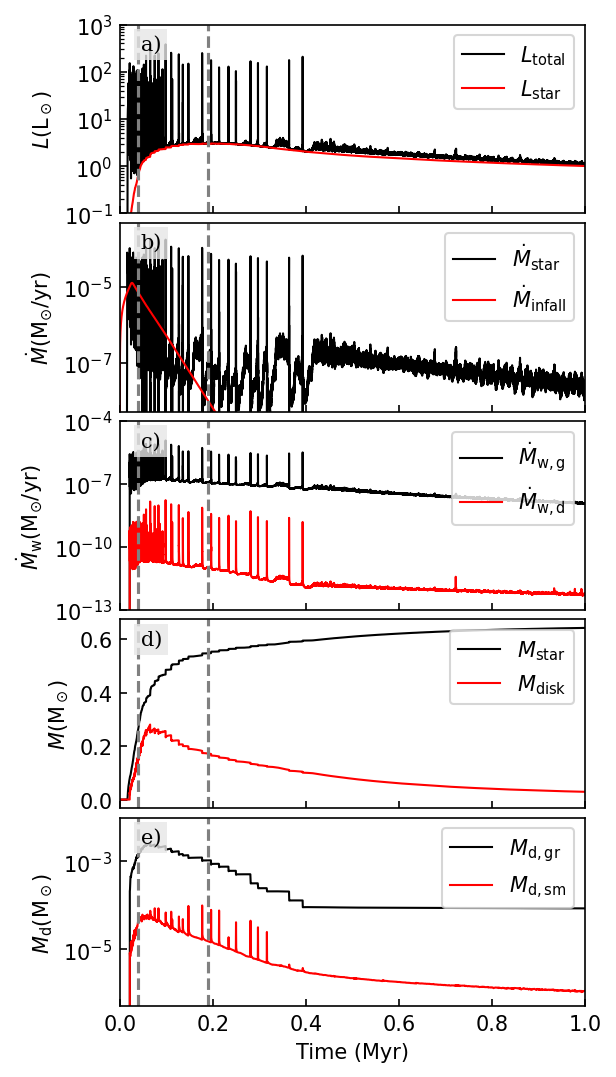}
\caption{Temporal evolution of the following global quantities over 1 Myr: 
a) total ($L_{\rm total}$) and stellar ($L_{\rm star}$) luminosity; 
b) accretion rate onto the star ($\dot{M}_{\rm star}$) and envelope infall rate ($\dot{M}_{\rm infall}$),;    
c) total wind mass-loss rate in gas ($\dot{M}_{\rm w, g}$) and dust ($\dot{M}_{\rm w, d}$);
d) stellar ($M_{\rm star}$) and disk ($M_{\rm disk}$) mass; and
e) total grown ($M_{\rm d,gr}$) and small ($M_{\rm d, sm}$) dust mass.
The vertical dashed lines depict the estimated class boundaries between Class 0/I and I/II.}
\label{fig:tGlobal}
\end{figure}

The particular simulation discussed in this section describes evolution of a PPD around a low-mass star and is identical to model WI-3a in \citet[][]{Kadam+25a}.
The mass of the initial cloud core is 0.83 $M_{\odot}$, while the maximum $\alpha_{\rm MRI}$ corresponding to a fully active disk is $10^{-3}$. 
The disk wind parameters considered offer a reasonable fit to the observations in terms of global disk properties such as the disk mass and radius.
Figure \ref{fig:tGlobal} shows the temporal evolution of the disk in terms of the characteristic global quantities over the 1 Myr of evolutionary period, starting from the beginning of the cloud core collapse.
The Class 0/I boundary (0.036 Myr) is defined as the time when the mass of the star-disk system exceeds half of the mass of the
initial cloud core, while the Class I/II boundary (0.187 Myr) is considered to be the point at which the envelope accretion rate falls below $10^{-8} {\rm M_\odot yr^{-1}}$.
The first panel shows the stellar photospheric luminosity and total (stellar+accretion) luminosity, while the second panel shows accretion rate onto the protostar as well as the cloud core infall rate onto the disk.
The accretion luminosity is punctuated by quasi-periodic and luminous outbursts of approximately 100 $L_\odot$, which correspond to accretion rates of $\sim 10^{-4} {\rm M_\odot/yr}$ as seen in panel b).
The accretion at early times is dominated by the outbursts, which continues into the Class II stage.
These correspond to MRI-type outbursts, which are triggered in the inner disk by an instability when viscous heating becomes sufficient to thermally ionize the alkali metals, which increases the degree of ionization -- and, hence, increasing $\alpha$  and causing amplified heating \citep{Bae+13,Kadam+20,VMHD20}. Such outbursts robustly occur in the early stages of the disk evolution; for example, for a wide range of stellar masses \citep{Elbakyan+21} and low metallicity environments \citep{Kadam+21}.

Panel c in Fig. \ref{fig:tGlobal} shows the gas and dust mass-loss rates of the magnetic wind ($\dot{M}_{\rm w,g}$ and $\dot{M}_{\rm w,d}$), integrated over the extent of the centrifugal disk.
During quiescence, $\dot{M}_{\rm w,g}$ is approximately equal to the stellar accretion rate, $\dot{M}_{\rm star}$.
This is partly achieved in our disk wind model via a multiplicative factor in the mass-loss rate (parameter $C_{H}$), which is not well constrained in shearing box simulations \citep[see][for details]{Kadam+25a}.
The atomic and molecular lines that trace slower and wider disk winds in the LVC component indicate a similar ratio of $\dot{M}_{\rm w,g}/\dot{M}_{\rm star}$ on the order of unity \citep{Pascucci+23}.
During an outburst, $\dot{M}_{\rm w,g}$ is about one order of magnitude less than $\dot{M}_{\rm star}$.
This occurs in the disk wind model through the dependence of the disk wind on the total luminosity of the central object, the latter being a proxy for the FUV irradiation of the disk.
Panel d shows the mass of the star as well as that of the disk.
The small discontinuities in both the masses correspond to the episodic accretion,  when a significant amount of mass (approximately a percent of the central star) is accreted onto the star. 
At 1 Myr, the disk mass is 0.05 ${\rm M_\odot}$, which is almost 10\% that of the central star. Thus, at 1 Myr, the system displays properties of a typical Class II TTauri star.
We emphasize that the number and magnitude of the outbursts is congruent with observations of outbursting YSOs \citep[e.g.,][]{Audard+14}, although the burst statistics have yet to be compared in detail.

The last panel e of Fig. \ref{fig:tGlobal} shows the grown and small dust mass contained within the entire disk.
The most salient feature of this plot is that (coincident with the outbursts) the grown dust content of the disk suddenly decreases, while the small dust content peaks.
We note that all peaks are not captured in the output, as the dust mass content of the disk is derived from 2D output files with a lower temporal resolution. 
The grown dust is prone to radial drift and, thus, it accumulates in the pressure maxima which are formed in the gas rings near the inner edge of the dead zone at a few au from the star \citep[see][for details on formation of rings]{Kadam+19,Kadam+22}.
During an MRI-type burst, the large grains are destroyed due to both increase in temperature and local $\alpha$, since the fragmentation size in inversely proportional to these quantities \citep{Birnstiel+12}.
Thus, large dust is converted into small dust, resulting in sudden increase in the small dust content. As the small dust component is well-coupled with the gas, it is accreted onto the star as the outburst proceeds, bringing the small dust content close to its original baseline.

\subsection{Outburst mechanism and dusty winds}

\begin{figure}
\centering
  \includegraphics[width=0.46\textwidth]{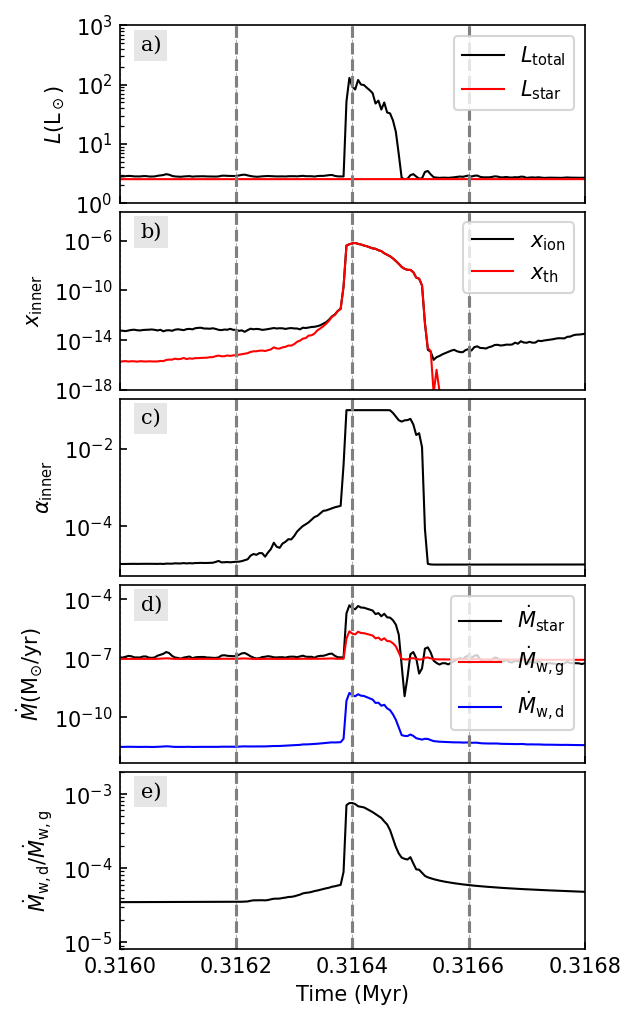}
\caption{Temporal evolution of some of the relevant quantities during an outburst event: 
{a) total and stellar luminosity; 
b) total ($x_{\rm ion}$) and thermal ($x_{\rm th}$) ionization fraction at the inner boundary;
c) $\alpha$-parameter at the inner boundary;
d) accretion rate onto the star ($\dot{M}_{\rm star}$); and total wind mass-loss rate in gas ($\dot{M}_{\rm w, g}$) and dust ($\dot{M}_{\rm w, d}$); and 
e) ratio of total dust to gas mass-loss rate of the wind.
} 
Vertical lines depict time slices which are expanded in terms of radial profiles in Fig. \ref{fig:rad}.}
\label{fig:tZoom}
\end{figure}

In Fig. \ref{fig:tZoom}, we zoom onto one particular burst occurring at approximately 0.3 Myr.
The duration of the luminosity burst in panel a is $\approx 100$ years, which is very well matched with canonical estimates of observed FU Ori events in YSOs \citep{Audard+14,Fischer+23}.
The second panel shows the total ($x_{\rm ion}$) as well as thermal ($x_{\rm th}$) ionization fractions, as measured at the inner boundary of the computational domain, at 0.51 au.
The total ionization fraction includes contribution from cosmic rays as well as short-lived radionuclides. 
As both of these quantities are measured at the inner boundary, they show a marginally longer duration of outburst, as compared to the luminosity.  
We note that the degree of thermal ionization starts rising before the luminosity outburst.  
In our current model, the value of $\alpha$ in the disk increases to the outbursting value of 0.1 above a threshold value of $x_{\rm th} = 10^{-10}$.
During the outburst, $x_{\rm th}$ rises steeply to maximum values that are consistent with total ionization of sodium at solar abundance \citep{Balduin+23}, dominating the total ionization fraction, $x_{\rm ion}$. 
Thus, we find that the results are not sensitive to the exact value of the chosen threshold in the model. 
The third panel of this figure shows $\alpha$ parameter at the inner boundary. It starts rising slowly before the outburst, since the ionization balance equation results in a progressively larger active layer thickness. Once the threshold value of $x_{\rm th} = 10^{-10}$ is exceeded, $\alpha$ achieves the maximum outbursting value of 0.1.

\begin{figure*}
\centering
  \includegraphics[width=0.95\textwidth]{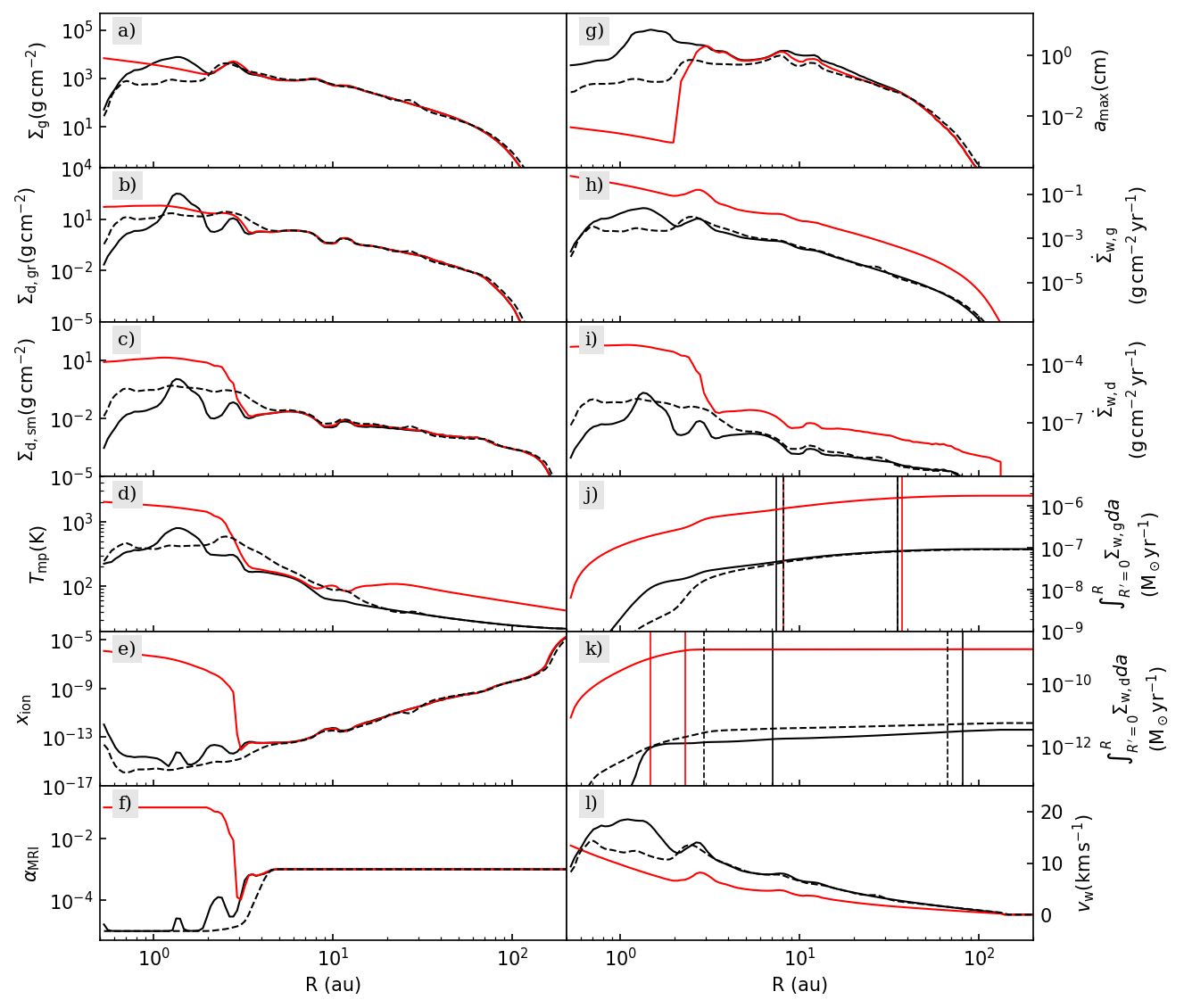}
\caption{Evolution of some of the azimuthally averaged quantities across an outburst, showing before (solid black), during (solid red) and after (dashed black) profiles. The selected phases for this plot are marked in Fig.~2.
The vertical lines in panels j and k correspond to the region which contains 50\% and 90\% of the total cumulative mass-loss rates, with the same color coding. }
\label{fig:rad}
\end{figure*}

Panel d of Fig. \ref{fig:tZoom} shows the mass accretion rate onto the star ($\dot{M}_{\rm star}$), as well as the mass ejection rates in both dust $\dot{M}_{\rm w,d}$ and gas $\dot{M}_{\rm w,g}$ due to disk winds, integrated over the centrifugal disk.
{Again, it can  be  clearly seen that the wind gas mass-loss rate is roughly equal to the accretion rate onto the star at quiescence, 
while during the outburst, this rate is about ten times smaller.} 
{The disk ejects more dust during an outburst as compared to that in quiescence.
This can be observed in the fourth panel, where the ratio of dust to gas mass-loss rate in the wind ($\dot{M}_{\rm w,d}/\dot{M}_{\rm w,g}$) is a proxy for the wind dust-to-gas ratio. 
During quiescence, the winds carry much less than the canonical ISM value (dust-to-gas ratio 0.01) of dust, as most of the dust mass resides as grown dust.
However, during an outburst, $\dot{M}_{\rm w,d}/\dot{M}_{\rm w,g}$ increases by over an order of magnitude, since a significant fraction of the grown dust is converted into small dust (see last panel of Fig. \ref{fig:tGlobal}).
Thus, our model predicts significantly more dusty winds during an outburst, which should eventually show as an extended dust emission beyond the disk.
{The increased production of small dust during an outburst can affect the ionization degree and in particular, the gas column density of the MRI-active region depends strongly on the ionization fraction of electrons.
This is taken into account in the ionization balance equation by considering the rate of recombination onto the two dust populations of dust grains \citep{VMHD20}.}
The last panel (e) of Fig. \ref{fig:tZoom} implies that quiescent winds are extremely dust-depleted, which is consistent with our understanding of disk outflows \citep{Suzuki+10,Gorti+15}.
Although the disk winds during outbursts are "dusty" compared to those in quiescence, they are still dust-depleted as compared to canonical ISM value.
The total amount of small dust ejected during a single event is typically of the order of $10^{-8} {\rm M_\odot}$. 
This is a very small fraction of the dust reservoir available for planet formation, as the parent cloud core of mass 0.83 ${\rm M_\odot}$ contains 1\% of dust.}

Figure \ref{fig:rad} shows the progression of the outburst in terms of some key azimuthally averaged quantities. Here, we aim to characterize the burst mechanism as well as the dust in the winds during an outburst by showing the structural evolution.
The three instances of time shown here correspond to the three vertical dashed lines plotted in Fig. \ref{fig:tZoom}, which represent the disk before, during (near the luminosity peak), and after an outburst.
The first three panels depict the change in the disk structure across the outburst. 
Panel a shows the gas surface density, which shows that the gaseous ring at $\approx$ 1.5 au is fully accreted onto the central star.
Panel b shows the destruction of grown dust during the outburst, as most of the dust gathered in the pressure maximum at $\approx$ 1.5 au in the outbursting region shows flattening of the surface density profile.
As most of the dust mass in a given cell is contained within the grown dust, its destruction leads to a significant increase in the small dust component.
We see this in the panel c, where the small dust surface density increases by over two orders of magnitude in the outbursting region, which extends approximately 3 au.

The next four panels of Fig. \ref{fig:rad} focus on the mechanism of the outburst and production of small dust.
The MRI-type outbursts occur naturally in our disk simulations and are already studied in detail \citep{Kadam+20,VMHD20}. 
In summary, the small viscosity in the dead zone is sufficient to increase the temperature in the innermost region. When it reaches the ionization temperature of alkali metals, the ionization fraction increases exponentially and the affected region becomes prone to MRI instability, thus triggering an outburst.
The increase in the midplane temperature during an outburst is seen in Panel d, while resulting increase in the thermal ionization fraction is seen in Panel e.
The resulting increase in the disk $\alpha$ to the maximum outbursting value of 0.1 is seen in Panel f.
Thus, we show that the fundamental mechanism of the outburst remains unchanged with our current model of $\alpha$ as a function of ionization fraction.
Both the increase in temperature and $\alpha$ during the outburst contribute towards a significant decrease in the fragmentation size of the dust particles.
In the inner disk, the dust growth is fragmentation-limited and thus, controls the maximum size of the grown dust component ($a_{\rm max}$).
As seen in panel g), $a_{\rm max}$ decreases by nearly three orders of magnitude during the burst in the inner disk.
This essentially causes the aforementioned destruction of the grown dust component. The small dust is thus suddenly available for in large quantity in the outbursting region of the inner disk.

The next four panels of Fig. \ref{fig:rad} focus on the properties of the ejected winds.
Panels h and i show the surface density loss rate due to winds for the gas and dust, respectively. 
Near the peak of the outburst, $\Sigma_{\rm w, g}$ increases throughout the disk, as per its dependence on the luminosity of the central star.
The dust loss rate shows a similar increase in the outer disk, however, $\Sigma_{\rm w, d}$ in the outbursting region increases significantly.
This is a result of the sudden production of the small dust, which is available to be ejected via winds.
Panels j and k show the cumulative mass-loss rate ($\int_{R'=0}^R \Sigma(R') da(R')$, where $da$ is the area element) due to gas and dust, respectively.
As expected, both of these quantities increase significantly during the outburst.
We go on to consider the vertical lines in these plots, which show the region enclosing 50\% and 90\% of the cumulative mass-loss rate. 
For the gas in the disk wind, the region within which a certain percentage of the ejected gas mass is contained remains relatively unchanged.
In case of the dust component, however, the mass-loss rate during an outburst is highly centrally concentrated, with 90\% of it coming from $< 3$ au of the inner disk. 
This reflects the ejection of large amount of dust mass from the affected innermost region during an outburst.

Lastly, as seen in panel l, the wind velocities are of the order of 10 of ${\rm km s^{-1}}$, which are consistent with observations of LVC components of disk winds \citep{Natta+14,McGinnis+18,Fang+23}. 
The outwards decreasing profile can possibly produce nested onion-like morphology of resolved sources \citep{Pascucci+23}.
We make a rough estimate of the wind velocities by equating the work done by the wind torque to the kinetic energy gained by the mass ejected from the disk (see appendix A of \cite{Kadam+25a}).
During the outburst, the wind velocity does not increase, but rather decreases by approximately a factor of two. 
This is because the enhanced wind carries a relatively large amount of mass as compared to wind torques, which results in a lower kinetic energy per unit mass, in turn, lowering the escape velocity.
This decrease is analogous to the anti-correlation between the ejected mass and the associated velocity of the winds during coronal mass ejection events \citep{CME11}.
Although the wind velocity is estimated in a naive fashion, for instance, without taking into account thermal effects or wind interactions, the available observations on outflow forces also support the idea of very low wind velocities during FUor outbursts \citep{Fernando+23}.

\section{Conclusions and discussions}
\label{sec:conclusions}

In this paper, we presented results from our numerical experiments of modeling the formation and evolution of PPDs, with a focus on episodic accretion and ejection of dust during such outbursts via magnetic winds.
The viscous $\alpha$ in the inner disk during the outbursts is modeled as a function of the thermal ionization fraction.
The disk winds are included with the formulation of \citet[][]{Kadam+25a}, wherein during the outbursts, both the wind mass-loss rate and torque increase as a function of the accretion luminosity.
We have shown the mechanism of the outbursts in detail, which is identical to MRI-type bursts and the burst properties remain unaltered across evolutionary time.
The qualitative agreement between an individual MRI-type eruption against observed longer duration FU Ori outbursts is excellent. 
This includes a fast rise time of $<10$ yr, slow decline with a total duration of $\approx$ 100 yr, luminosity of few hundred $L_\odot$, and peak accretion rate of $10^{-4} {\rm M_\odot yr^{-1}}$

The MRI-type outbursts arise from residual viscous heating in the dead zone, which gradually results in midplane temperatures high enough to thermally ionize alkali metals.
This leads to a sudden increase in the ionization fraction and thus effective $\alpha$, as the disk becomes MRI-active. 
During the outburst, the increase of temperature and $\alpha$ in the affected inner disk region result in a lowering of the dust fragmentation size and, hence, the destruction of the grown dust.
The additional small grains suddenly become available for ejection with the magnetic winds, as only the small particles are aerodynamically coupled and entrained in the winds.
One caveat of the model is that dust is considered to be formed of icy aggregates (small silicate monomers held together by water ice), which are destroyed instantaneously with an increase in temperature. 
However, such dust particles may be resilient to destruction at the onset of accretion events \citep{Houge+24}; hence, we plan to consider a ``slow'' collisional fragmentation model in the future to confirm the robustness of our findings.
Another limitation of our simulations is that they are performed in two dimensions with thin-disk approximation and we have no information of the winds once they leave the computational domain. The extension of our model into the third dimension to track the disk winds remains a task for future exploration.

During an outburst, the dust mass-loss rate integrated over the disk increases by almost two orders of magnitude, as compared to the quiescent state.
However, the winds during outburst are still about an order of magnitude dust-depleted as compared to the ISM.
The innermost regions affected by the outburst contribute most to the dust mass-loss rate, with 90\% of the dust coming from innermost few au.
The quantity of dust ejected does not significantly affect the mass reservoir available for planet formation. 
Since direct observations of outflows tracing dust emissions during FUor outbursts are rare, a concrete comparison with observations is an objective to be addressed in future work.
However, our model of disk winds and resulting outflows during eruptions is in alignment with several observations associated with YSOs and young eruptive stars.
For example, a wind mass-loss rate matching the central accretion rate during quiescence, wind mass-loss rate of approximately 10\% during eruption, dust-depleted winds during quiescence, entrainment of sub-micron-sized dust by the winds, decreased velocity of outflows during eruptions, the possibility of nested onion-like morphology with outwards decreasing velocity profile, and, finally, increased dust content in the winds during eruptions. 
Our study sheds light on the mechanism of dust production during outbursts and makes a predication of an order of magnitude  on the dust mass loss rates in outflows and its origin in the innermost regions.
Dusty winds during outbursts are expected to exhibit characteristic signatures, such as a periodic increase in the dust continuum emission of the system.

\begin{acknowledgements}
We thank the anonymous referee for constructive feedback, which improved the quality of this article.
The numerical simulations were performed with High-Performance Computation support provided by the Digital Research Alliance of Canada.
E. I. V. acknowledges support from the Ministry of Science and Higher Education of the Russian Federation (State assignment in the field of scientific activity 2023, GZ0110/23-10-IF).
This research has made use of the Astrophysics Data System, funded by NASA under Cooperative Agreement 80NSSC21M00561.
\end{acknowledgements}

%
%

\bibliographystyle{aa}
\bibliography{references}

\end{document}